\documentclass[twocolumn,twoside,slac_two]{revtex4}
\usepackage{graphicx}
\usepackage{fancyhdr}
\usepackage{graphics}
\usepackage{epstopdf}
\usepackage{textpos}
\def \invfb{ fb\ensuremath{^{-1}} }

\begin{document}

\title{\centering Exotic Searches at LHC and Tevatron}


\author{
\centering
\includegraphics[scale=0.15]{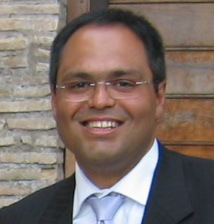} \\
\begin{center}
Sh.~Rahatlou, on behalf of ATLAS, CDF, CMS, and D0 Collaborations
\end{center}}
\affiliation{\centering Sapienza Universit\`a di Roma \& INFN, Rome, 00185, Italy}
\begin{abstract}
Successful operation of the Large Hadron Collider has led to more than 1\invfb of data 
recorded with both ATLAS and CMS detectors by summer of 2011. This large amount of data has allowed to perform
numerous searches for rare processes beyond the Standard Model, many of which are competitive with 
previous searches performed with the CDF and D0 detectors at the Tevatron. In this talk the most recent searches
at the hadron colliders are reviewed.
\end{abstract}

\maketitle
\thispagestyle{fancy}


\section{INTRODUCTION}

Searches for exotic processes of New Physics beyond the Standard Model have been a major focus
of analysis efforts in the last few years at the Tevatron. The successful operation of the Large Hadron Collider (LHC) in 2010
at a center-of-mass energy of $\sqrt{s} = 7$~TeV opened a new season in the exotic searches, providing a new opportunity to spot
signatures of new physics. The cross section of rare processes profits greatly from the higher center of mass energy and hence 
the very first searches at the LHC with just 35~$pb^{-1}$ of accumulated data in 2010 could compete immediately with the results 
obtained with the CDF~\cite{CDF} and D0~\cite{D0} detectors at the Tevatron, operating at $\sqrt{s} = 1.96$~TeV. 
The performance of the LHC in 2011 by far exceeded even the more optimistic
predictions and both ATLAS~\cite{ATLAS} and CMS~\cite{CMS} detectors recorded more than 1\invfb of data before Summer. This thirty-fold increase
in recorded luminosity combined with the excellent performance of the Tevatron in the same period led to more than 60 new results
from the four collaborations involved in the exotic searches. Due to time and space constraints, this article cannot serve as a 
comprehensive review of all ongoing efforts but rather as a snapshot of the latest results using the largest available data samples at
the LHC and Tevatron. A complete list of all public results from the four collaborations, together with additional supporting material not included 
in the papers or conference notes, is available on the web for ATLAS~\cite{atlas:pubexo}, CDF~\cite{cdf:pubexo}, CMS~\cite{cms:pubexo}, and
D0~\cite{d0:pubexo}.
Another challenge in reviewing such a large number of results, is grouping them in coherent and related groups. 
A given experimental signature, e.g. di-lepton final state, is predicted by several theoretical models, including new symmetries 
requiring new heavy bosons and  theories of extra dimensions. If final states with two, three, of four objects are considered, 
where each object can be a jet, lepton, photon, or missing transverse energy (MET), the number of final states is limited and well defined.
Rather than using a signature-based approach, in this review searches are grouped in four topics of interest: 
heavy resonances, both leptonic and hadronic,
large extra dimensions,
long-lived particles, 
and $4^{th}$ generation quarks.
For searches with a top quark or $t\bar{t}$ in the final state see the proceedings of this conference~\cite{spano}. For SUSY searches 
see the proceedings of this conference~\cite{portell}.

\section{HEAVY RESONANCES}
Many models of new physics predict the existence of narrow resonances, possibly at the TeV mass scale, that decay to a pair of charged leptons.
Some of the more popular models used as benchmark reference include 
the Sequential Standard Model $Z^\prime_{SSM}$ with standard-model-like couplings, the $Z^\prime_\psi$ predicted
by grand unified theories ~\cite{GUT}, and Kaluza--Klein graviton excitations arising in the Randall--Sundrum (RS) 
model of extra dimensions ~\cite{RS1,RS2}.

The results of searches for narrow $Z^\prime \rightarrow l^+l^{-}$ and $G_{KK} \rightarrow l^+l^{-}$ resonances in $p\bar{p}$ 
collisions at the Tevatron with over 5~fb$^{-1}$ of integrated luminosity at centre-of-mass energy of $1.96$ TeV
have previously been published~\cite{D0RS,D0Zprime,CDFll1,CDFll2}. The most recent unpublished spectrum from the CDF collaboration at Fermilab 
in the dielectron channel~\cite{CDFll3}  has three events above 600 GeV, the highest at 960 GeV, with an integrated luminosity of 5.7fb$^{-1}$. 

Both ATLAS and CMS collaborations have searched for such narrow resonances in, respectively, 1.2fb$^{-1}$ and 1.1fb$^{-1}$ of data collected at the LHC 
until the summer of 2011 in  dilepton (both electrons and muons)~\cite{ATLASll,CMSll} and diphoton~\cite{ATLASgg,CMSgg} final states. 
The observed invariant mass spectra agree with expectations based on standard  model processes. Therefore limits are set on the mass of a narrow heavy resonance. 
Summary of the observed limits is reported for dileptons in Table~\ref{tab:ll} and for diphoton final state in table~\ref{tab:gg}.
\begin{table}[t]
\begin{center}
\caption{Summary of exclusion limits for the mass of a narrow resonance decaying to dilepton final states. All  values are in TeV and $k$ is the curvature of the extra dimension
in the RS model.}
\begin{tabular}{|l|c|c|c|c|}
\hline
 & $Z^\prime_{SSM}$ & $Z^\prime_{\psi}$ & $G_{KK}$, $k=0.05$ & $G_{KK}$, $k=0.10$ \\
\hline
ATLAS & 1.83 & 1.49 & 1.33 & 1.63 \\
CMS & 2.00 & 1.62 & 1.49 & 1.79 \\
\hline
\end{tabular}
\label{tab:ll}
\end{center}
\end{table}
\begin{table}[t]
\begin{center}
\caption{Summary of exclusion limits for the RS Graviton in the diphoton final state. All values are in MeV and $k$ is the curvature of the extra dimension
in the RS model.}
\begin{tabular}{|l|c|c|c|c|}
\hline
  & CMS  & ATLAS & CDF ($ee+\gamma\gamma$) & D0 ($ee+\gamma\gamma$) \\
  & 1.1fb$^{-1}$ & 36pb$^{-1}$ & 5.7fb$^{-1}$ & 5.4fb$^{-1}$ \\
\hline
$k=0.05$ & 1360 & 700 & 937 & 940 \\
$k=0.10$ & 1685 & -- & 1055 & 1050 \\
\hline
\end{tabular}
\label{tab:gg}
\end{center}
\end{table}

Searches have been conducted also for resonances in the dijet final states. No excess of events has been observed beyond the expected background from
the Standard Model and limits have been set on the mass on such heavy resonances for a variety of reference models. The most stringent
limits today are those obtained by the ATLAS~\cite{dijet:ATLAS} and CMS~\cite{dijet:CMS} and are summarized in table~\ref{tab:dijet} for a few reference models.
\begin{table}[t]
\begin{center}
\caption{Summary of exclusion limits for a narrow dijet resonance. All values are in TeV.}
\begin{tabular}{|l|c|c|c|c|}
\hline  
  & $q^*$                                          & Axigluon~\cite{axigluon} & color octet                       & string reso-\\
  &     \cite{qstar1,qstar2}                  &  coloron~\cite{coloron} &  scalar ~\cite{coloctet}       & nances~\cite{string1,string2} \\
\hline
ATLAS & 2.99 & 3.32 & 1.92 &   --   \\
CMS    & 2.49 & 2.47 & --     &  4.00 \\
\hline
\end{tabular}
\label{tab:dijet}
\end{center}
\end{table}

Heavy W-like resonances are also predicted by several extensions of the Standard Model. In one such benchmark model~\cite{wprime}, 
the W$^\prime$ boson is considered a heavy analogue of the SM W boson with the same left-handed fermionic couplings. 
Interactions of the W$^\prime$ boson with the SM gauge bosons and other heavy gauge bosons such as the Z$^\prime$ are excluded.
However, in some models coupling to the leptons is suppressed, leading to a relative enhancement in the triple gauge couplings that could 
lead to a WZ final state~\cite{WZ}.
Searches for heavy W$^\prime$ bosons have been performed at the Tevatron~\cite{WZ:D0} for many years and started at the LHC since 2010 using leptonic final states.
The most stringent lower limit on the W$^\prime$ mass is about 2.3 TeV and has been obtained at ATLAS~\cite{wprime:ATLAS} and CMS~\cite{wprime:CMS}.
In the WZ final state masses below 784~GeV have been excluded~\cite{wprime:WZ}, 
assuming the extended gauge model for the coupling of W$^\prime$ to WZ.

The search in the WZ final state is also interpreted in the context of 
Technicolor (TC), a strongly interacting gauge theory which allows for the dynamical breakdown of electroweak symmetry [12, 13]. The lightest 
$\rho_{TC}$ and $\omega_{TC}$ are expected to have masses below  $\sim$700 GeV, and their decay channels (e.g. $\rho_{TC} \rightarrow WZ$) 
have distinctive signatures with narrow resonant peaks. A $\rho_{TC}$ with a mass below 382 GeV in the parameter space $M(\pi_{TC}) = 3 M(\rho_{TC}) - 25$ 
GeV has been excluded, as well as $\rho_{TC}$Õs with masses below 436 GeV in the parameter space $M(\rho_{TC}) < M(\pi_{TC}) + M_W$
~\cite{wprime:CMS}. 
These are the strongest limits to date in this channel.

Recently the CDF collaboration reported an excess in the dijet mass spectrum at 145 GeV~\cite{wjj:cdf1} in events with two jets produced 
in association with a W boson. An updated search with 7.3fb$^{-1}$ at CDF~\cite{wjj:cdf2} confirms this excess which however is not confirmed
by the D0 collaboration with 5.4fb$^{-1}$ of data~\cite{wjj:d0}. A similar search by the ATLAS collaboration at the LHC also does not observe any excess of events
beyond the Standard Model bakground~\cite{wjj:atlas}.

Finally, searches have been conducted by the ATLAS~\cite{lljj:atlas} and CMS~\cite{lljj:cms} collaborations in the $lljj$ final state sensitive 
to the existence of a heavy neutrino and a right-handed heavy W-like boson
as predicted in left-right (LR) symmetric extensions to the Standard Model model~\cite{lljj:1,lljj:2,lljj:3}, which naturally explain the parity violation 
seen in weak interactions as a result of spontaneously broken parity. No excess of events is observed beyond the expected Standard Model background
and exclusion limits are set as a function of the heavy-neutrino and right-handed W$_R$ masses. 
For a heavy neutrino with masses up to 1 TeV, 
the exclusion contour extends to W$_R$ masses of up to 1.6 TeV in both electron and muons channels.

\section{EXTRA DIMENSIONS}

The existence of extra spatial dimensions is an intriguing scenario that may solve the hierarchy problem~\cite{hierarchy} 
of the Standard Model. The original proposal to use extra dimensions (ED) to solve the hierarchy problem 
was presented by Arkani-Hamed, Dimopoulos, and Dvali (ADD)~\cite{ADD1,ADD2,ADD3}. They posited a scenario wherein the SM 
is constrained to the common 3+1 space-time dimensions (brane), while gravity is free to propagate through the 
entire multidimensional space (bulk). Thus, the gravitational flux in 3+1 dimensions is effectively diluted by virtue of 
the multidimensional Gauss's Law. The fundamental Planck scale $M_D$ is therefore related to the apparent scale $M_{Pl}$
 according to the formula  $M_{D}^{n_{ED}+2} = M_{Pl}^2 / r^{n_{ED}}$	where $r$ and $n_{ED}$ are the size and 
 number of the EDs, respectively.
 The coupling of the Kaluza-Klein modes to the SM energy-momentum tensor results in an effective theory with virtual 
 graviton exchange at leading order in the perturbation theory. A phenomenological consequence is a non-resonant 
 enhancement of expected dilepton and diphoton events at high invariant masses. The CMS collaboration has searched for such 
 an  excess of events in the invariant mass region  $> 0.8$~TeV in the diphoton final state~\cite{CMSgg} and $>1.1$~TeV in the dimuon
 final state~\cite{ADDmumu:CMS}. No events in excess of the expected  SM background, dominated respectively by QCD and Drell-Yan, are found
 and exclusion limits are set in the parameter space of the model for $n_{ED}=2,3,4,5,6,7$.
 
 Other direct signatures of ADD include the direct production of gravitons  in association with one energetic jet or photon. 
 The gravitons are very weakly coupled and their presence is inferred from the missing transverse energy $E_T^{miss}$.
 The primary SM backgrounds in these channels are the $W/Z+$jets events with the Z decaying in the invisible channel $\nu\bar{\nu}$ and
 the W decaying leptonically. In searches for these signature by the ATLAS and CMS 
 collaborations~\cite{monojet:ATLAS, monojet:CMS,monophoton:CMS} the data are found to be in agreement 
 with the expected contributions from SM processes and exclusion limits are set on $M_D$ for different values of $n_{ED}$ which  
 significantly improve the previous limits for this model from previous searches at LEP, Tevatron, and LHC.


\section{LONG-LIVED PARTICLES}

Many extensions of the standard model predict the existence of new Heavy stable or quasi-Stable Charged Particles~\cite{HSCP1} (HSCP). 
Such particles are present in some supersymmetric models~\cite{HSCP2,HSCP3,HSCP4}, and  are 
also a hallmark of split supersymmetry~\cite{HSCP5}, where the gluino ($\tilde{g}$) decay is suppressed due to the large gluino-squark mass splitting, 
from which the theory gets its name.  If long-lived gluinos (stops) are produced at the LHC, they will hadronise into 
$g\tilde{g}$, $\tilde{g}q\bar{q}$, $\tilde{g}qqq$ ($\tilde{t}q, \tilde{t}q\bar{q}$) 
states which are collectively known as ÒR-hadronsÓ.
If the lifetime of an HSCP produced at the Large Hadron Collider (LHC) is longer than a few nanoseconds, the particle will travel over 
distances that are comparable or larger than the size of a typical particle detector. In addition, if the HSCP mass is $\ge100$~GeV, 
a significant fraction of these particles will have a velocity, $\beta = v/c$, smaller than 0.9. These HSCPs will be directly observable 
through the distinctive signature of a high momentum particle with an anomalously large rate of energy loss through ionization 
$dE/dx$ and an anomalously long time-of-flight (TOF). For low-$\beta$ R-hadrons, this energy loss is sufficient to bring a significant fraction of the 
produced particles to rest inside the detector volume~\cite{HSCP6}. 

No significant excess above background is observed in a search for long-lived particles which have stopped in the CMS detector after being 
produced in the pp collisions followed by the subsequent decay of these particles during time intervals where there were no pp collisions 
in CMS~\cite{EXO11020}.
An upper limit has been set on the cross section of the HSCP pair production over 13 orders of magnitude of HSCP lifetime.

For a mass difference $m_{\tilde{g}} - M_{\tilde{\chi}^0_1} > 100$ GeV, assuming BR($\tilde{g} \rightarrow g\tilde{\chi}^0_1$) = 100\%, 
a gluino with lifetimes from 10~$\mu s$ to 1000~s and 
$m_{\tilde{g}} < 601$ GeV is excluded. Under similar assumptions,
$m_{\tilde{t}} - M_{\tilde{\chi}^0_1} > 200$~GeV, and
 BR($\tilde{t} \rightarrow t\tilde{\chi}^0_1$) = 100\%, a stop with lifetimes
from 10~$\mu s$ to 1000~s and $m_{\tilde{t}} <  337$~GeV is excluded. 
This result is consistent with the complementary exclusion provided by the direct search for a charged 
highly ionizing particle through its interaction in the tracker and the muon detectors~\cite{EXO11022}. 
These results improve previous limits obtained at ATLAS~\cite{HSCP:ATLAS} and CMS~\cite{HSCP:CMS2010} with the
 2010 dataset. 

The D0 collaboration has recently performed a similar search which excludes pair-produced long-lived 
gaugino-like charginos below 267~GeV and higgsino-like charginos below 217 GeV~\cite{HSCP:D0}.

\section{$4^{\rm th}$ GENERATION}

Since the discovery of the top quark at the Tevatron, there have been many searches for a possible 
new generation of fermions. Those searches have not found evidence of new fermions beyond the standard model (SM). 
However, from a theoretical point of view, the number of generations of fermions is not limited to three. 
The extension of the generations of fermions may have a significant effect on neutrino physics, flavor physics and Higgs physics. 
With a fourth generation, indirect bounds on the Higgs boson mass can be relaxed~\cite{4gen1,4gen2}, and an additional generation of quarks 
may possess enough intrinsic matter and anti-matter asymmetry to be relevant for the baryon asymmetry of the Universe~\cite{4gen3}. 
Therefore, there is continued theoretical and experimental interest in such a fourth generation~\cite{4gen4}. 
Direct searches restrict the masses of quarks in the fourth generation, $t^\prime$ and $b^\prime$, to be greater than 
350 GeV$/c^2$~\cite{4gen5,4gen6} , 
and the indirect search from LEP excludes a fourth type of light neutrino~\cite{4gen7}. At the LHC, the QCD
production cross section of $t^\prime\bar{t}^\prime$ is expected to be significantly larger than that at the Tevatron~\cite{4gen8}.
This brings us a great opportunity to explore the possibility of new physics with an extended generation of fermions.
Several searches at both Tevatron and LHC are presented here which exploit the rich number of final states available following
the $t^\prime\bar{t}^\prime$ production. The main background for this searches is due to the Standard Model $t\bar{t}$  
and $W/Z+$jets events which could emulate the signal. However, requirements on the total visible transverse energy and the missing 
transverse energy are typically sufficient to achieve good background rejection and define signal regions with good $S/B$ ratio.
The typical decay chain considered in these searches is given by $t^\prime/b^\prime \rightarrow t/b + X$ where $X$ can be either a $W/Z$ boson or 
a new particle $X$ escaping detection. Therefore the specific final states can vary from all hadronic, when both the top and the $W/Z$ decay hadronically 
to leptonic only, when only the leptons are used. In all these searches, using different amount of data, no excess
is observed beyond the expected Standard Model background and lower limits are set on the mass of the fourth generation fermion. 
These limits are summarized in table~\ref{tab:tprime}.

\begin{table}[t]
\begin{center}
\caption{Summary of exclusion limits for a fourth generation quark.}
\begin{tabular}{|l|c|c|c|c|}
\hline  
\textbf{Decay} & \textbf{Exp.} & \textbf{Method} & \textbf{Excluded} & \textbf{Lum.}  \\ 
                       &     &      &     \textbf{mass} (GeV)   & (\invfb) \\
\hline
$b^\prime \rightarrow t+W$ & CMS~\cite{EXO11036} & lep+jets & 495 & 1.1 \\
$b^\prime \rightarrow t+W$ & CDF~\cite{bprime:CDF}& lep+jets & 371 & 4.8 \\
$Q_4 \rightarrow q+W$ & ATLAS~\cite{q4:ATLAS}& lep+jets & 270 & 0.035 \\
\hline
$t^\prime \rightarrow b+W$ & CMS~\cite{EXO11050}& dilepton & 422 & 1.1 \\
$t^\prime \rightarrow b+W$ & CMS~\cite{EXO11051}& lep+jets & 450 & 1.1 \\
$t^\prime \rightarrow b+W$ & CDF~\cite{CDF10395}& lep+jets & 358 & 5.6 \\
$t^\prime \rightarrow b+W$ & D0~\cite{tprime:D0}& lep+jets & 285 & 5.3 \\
\hline
$t^\prime \rightarrow t+Z$ & CMS~\cite{EXO11005} & lep+jets & 417 & 0.2 \\
\hline
$t^\prime \rightarrow t+A_0$ & ATLAS~\cite{ATLAS-CONF-2011-03}& dilepton & 290 & 0.034 \\
\hline
$t^\prime \rightarrow t+X$ & CDF~\cite{tXhad:CDF}& hadronic & 400 & 5.7 \\
$t^\prime \rightarrow t+X$ & CDF~\cite{tXlj:CDF}& lep+jets & 360 & 4.8 \\

\hline
\end{tabular}
\label{tab:tprime}
\end{center}
\end{table}

\section{CONCLUSIONS}
The outstanding performance of the LHC and the ATLAS and CMS detectors led to more than 1\invfb of
accumulated data by summer of  2011 exceeding the expectations. This large amount of data
has allowed a multitude of searches to be conducted in the first year of LHC operation, with results
are are competitive, and in many cases improving the existing limits from the Tevatron. In particular the 
data acquisition and the computing infrastructure for data delivery worked according to the design specifications
enabling thousands of users around the globe to have timely access to the flow of data delivered by the LHC.
The detector performance also exceeded the expectations, with the missing  transverse energy and the $b$-flavor
tagging techniques playing a crucial role in many of the searches presented here. Unfortunately no excess of events
has been observed and the measurements are in good agreement with the Standard Model expectations.



\bigskip 
\bibliography{basename of .bib file}

\end{document}